\documentclass[twocolumn,showpacs,preprintnumbers,amsmath,amssymb,pre]{revtex4}

\usepackage{graphicx,bm}

\begin{document}

\title{Scale-invariant geometric random graphs}

\author{Zheng Xie$^{1}$ and Tim Rogers$^{2}$}

\affiliation
{
	$^{1}$College of Science, National University of Defense Technology, Changsha, 410073,  China\\
	$^{2}$Centre for Networks and Collective Behaviour, Department of Mathematical Sciences, University of Bath, Bath,
	BA27AY, UK.
}

\begin{abstract}
We introduce and analyze a class of growing geometric random graphs that are invariant under rescaling of space and time. Directed connections between nodes are drawn according to influence zones that depend on node position in space and time, mimicking the heterogeneity and increased specialization found in growing networks. Through calculations and numerical simulations we explore the consequences of scale-invariance for geometric graphs generated this way. Our analysis reveals a dichotomy between scale-free and Poisson distributions of in- and out-degree, the existence of a random number of hub nodes, high clustering, and unusual percolation behaviour. These properties are similar to those of empirically observed web graphs.
\end{abstract}


\maketitle

\section{Introduction}

\begin{figure}[t]
\includegraphics[width=240pt, trim=0 0 0 0]{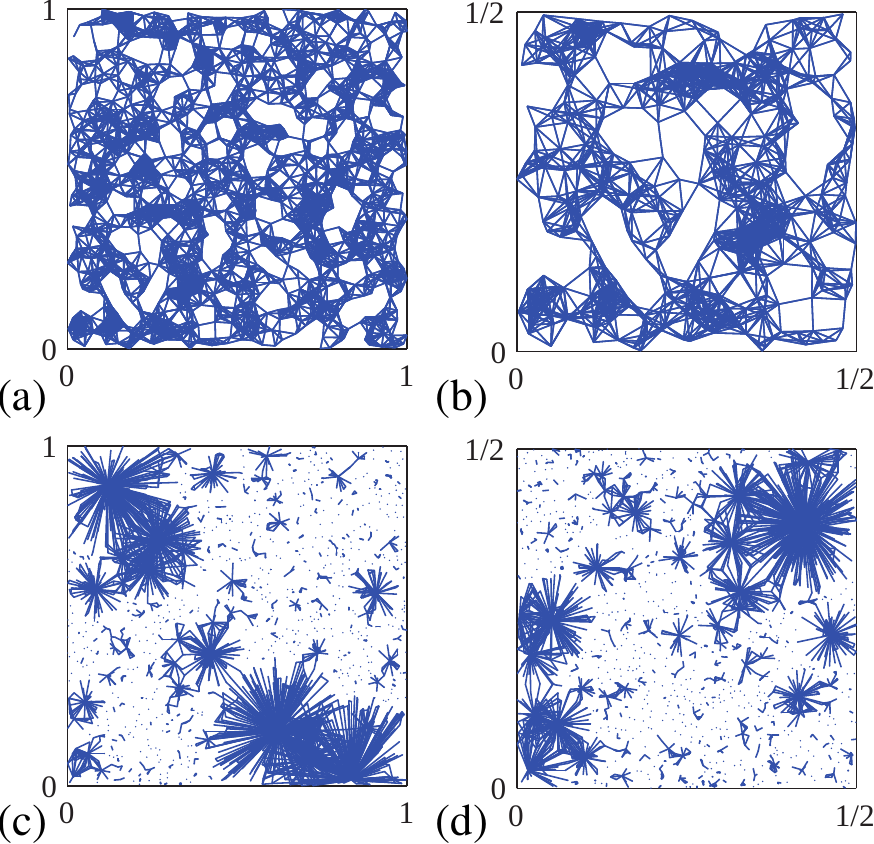}
\caption{Top row: an RGG in the domain $[0,1]^2$, with $N=1500$ nodes and fixed connection distance radius $d=0.05$, shown in full (a) and in close-up (b). Bottom row: a scale-invariant RGG in the same domain, with $N=10000$ nodes and influence zone areas $a_i=t_i^{-1}/2$, shown in full (c) and in close-up (d). This choice of network ensemble is a special case of the general form given in Eq.~(\ref{influ}).}
\label{fig1}
\end{figure}

The theory of random geometric graphs (RGG) enables research of complex networks via geometry \cite{Dall,Penrose,Barthelemy}. It is attractive to imagine the nodes of a complex network embedded in space, as this can provide useful intuition about an otherwise complicated and abstract discrete structure. The standard RGG can be formulated as follows: starting with an empty graph, new nodes arrive at a rate of one per unit time, they are placed at random location in the unit square, and attached to all existing nodes within distance $d$. The process is stopped when we reach a graph of size $N$. The properties of networks generated this way have been exhaustively studied, with particular emphasis on percolation \cite{Penrose,Alon,Ziff}. In applications, some specific networked systems are well-described by models of this type (\textit{e.g.} relations between mobile telephone users and cell sites \cite{Jayanth,Haenggi,Mukherjee}), and the propensity of new nodes to attach only to those that are `similar' (in the sense of spatial location) mirrors the homophily frequently observed in social systems \cite{Sim}. However, these model networks exhibit a fixed natural scale (see left panels of Fig.~\ref{fig1}) and a level of homogeneity that is quite atypical of many real world networks \cite{Newman2}.

The now textbook explanation for the extreme node inhomogeneity observed in many real networks is as a result of growth via a mechanism of preferential attachment \cite{Barabasi}. This idea has recently been generalized to a geometrical setting in the `popularity and similarity' models of web graph formation \cite{Menczer,Papadopoulos}. The consequences of preferential attachment for the network include a scale-free degree distribution and high clustering coefficient, however, these same effects can also be generated by other mechanisms. Recent work in so-called `network cosmology' considers the modification of random geometric graph attachment rules to take account of node `visibility' according to the light cone structure of some pseudo-Riemannian manifold (e.g., de Sitter space, the standard cosmological model \cite{Ahmed,Bombelli,Rideout}), which can result in a scale-free network \cite{Krioukov}. This approach is meaningful in the case of systems with an underlying causal structure, such as citation networks \cite{Xie}.

In this article we reverse the question: rather than asking what causal mechanisms might give rise to a scale-free network, we ask what are the consequences of demanding that a random geometric graph ensemble is invariant to its spatial and temporal scales? We consider a RGG in which nodes are created by a unit intensity Poisson process in a three-dimensional Euclidean volume (two spatial dimensions, plus a time axis) and connected according to a function of birth time and spatial proximity. The novel aspect we introduce is to demand that the statistics of the resulting graph ensemble depend only on total size and not the spatial or temporal scale (see the right panels of Fig.~\ref{fig1}). As we show, this simple constraint induces a range of emergent properties in the generated networks: a dichotomy between scale-free and Poisson distribution of in- and out-degree, the existence of a random number of hub nodes, high clustering, and unusual percolation behaviour. 

It turns out that, in this context, scale-invariance is synonymous with a particular form of increasing specialization of nodes, \textit{i.e.} that on average the connection radius is smaller for newer nodes. The increase of specialization \textit{in general} is a common feature of many real-world systems; examples of this phenomenon include the progression of scientific research, the replacement of generalist species by specialists in ecology, and the evolution of the world-wide-web. For this last example, one might think of nodes as web pages, with the spatial coordinates as describing position in some abstract `content space' (much like in the models of \cite{Menczer,Papadopoulos}), where newer pages typically have a smaller scope. Although this story is far from a perfect description of the true mechanisms behind the formation of web graphs, we will show that some of the important properties of the Stanford webgraph can be fit to a simple example of our scale-invariant random graphs. 

The paper is organised as follows. In Section~\ref{moddef} we present a general formulation geometric random graphs, and show how demanding scale-invariance places tight constraints on the model specification. The distributions of in- and out-degree for scale-invariant geometric random graphs are derived in Section~\ref{degs}, and Section~\ref{cluperc} presents some tentative results concerning clustering and percolation in these graphs. Finally, in Section~\ref{web} we briefly illustrate the agreement between certain properties of scale-invariant geometric random graphs and real-world webgraphs. 

\section{Model definition}\label{moddef}

\begin{figure}[t]
\includegraphics[width=0.4\textwidth, trim=0 0 0 0]{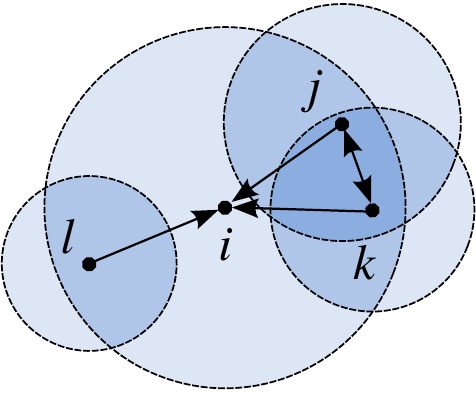}\
\caption{Illustration of the node attachment rule. In this example the nodes arrived in the time-order $i,j,k,l$ and have correspondingly ordered zonal areas.}
\label{ill}
\end{figure}

We consider the following general attachment rule for growing random geometric graphs. Nodes appear as a unit intensity Poisson process in time (\textit{i.e.} the time between one node appearing and the next is standard exponential random variable, independent of everything else in the system). Node $i$ is identified by its location in space $\bm{x}_i\in\mathbb{R}^2$ and time $t_i\in\mathbb{R}^+$. Each node is assigned a spatial zone of influence (or simply \emph{`zone'} hereafter) defined as the circular region with center $\bm{x}_i$ and area $a_i=f(\bm{x}_i,t_i)$, where $f$ is a positive and piecewise-continuous function. For each ordered pair of nodes $i$ and $j$, a directed edge is drawn $j\to i$ if $\bm{x}_j$ lies within the zone of $i$ (see Fig.~\ref{ill} for an illustration). Networks are sampled by restricting the model to a finite region $\Omega$; for simplicity we choose a cylinder with radius $R$ around the origin in space and duration $T$ in time. 

For given $R$ and $T$ the expected number of nodes is $\mathbb{E}N=\pi R^2T$. The standard RGG is represented in this class by the choice $f(\bm{x},t)=\pi d^2$. Note that varying $R$ and $T$ but keeping $\mathbb{E}N$ fixed will cause networks generated with the standard RGG to have very different statistical properties; if $R\gg d$ then the graph will have almost no edges, if $R<d$ then it will be almost fully connected. What happens if we demand that the law of the random graph depends only on $\mathbb{E}N$ and is insensitive to the choice of $R$ and $T$?

This demand ensures that $f$ must take a particular functional form. Considering the law of a single node, we have the requirement that for arbitrary scale factors $\lambda,\mu>0$
\begin{equation}
f(\lambda\bm{x},\mu t)=g(\lambda)h(\mu)f(\bm{x},t)\,,
\label{si}
\end{equation}
for some pair of functions $g$ and $h$. Using polar spatial coordinates $\bm{x}=(r,\theta)$, it is straightforward to check that the only positive continuous solutions to (\ref{si}) have the form $f(\bm{x},t)=\alpha(\theta)t^{-\beta}r^{-\gamma}$, where $\alpha$ is a bounded function and $\beta,\gamma$ are constants. Moreover, it turns out that consideration of the graphs as a whole provides a joint scaling relation that the exponents $\beta$ and $\gamma$ must satisfy. 

For a fixed expected number of nodes $\mathbb{E}N$, we require that the expected number of edges $\mathbb{E}M$ does not depend on the spatiotemporal scale. The total number of edges in a given graph is $M=N\langle k^-\rangle/2$, where $\langle k^-\rangle$ is the average in-degree of nodes in that graph. For now we are only interested in scaling behaviour in large graphs, so we assume that (by the law of large numbers) we may conflate the arithmetic average over the nodes of a particular random realization with an ensemble average. Since nodes fall as a unit rate Poisson process, the ensemble average degree of node $i$ is simply $a_iT$. Ignoring boundary effects and correlations between vertices, we can thus compute
\begin{equation}
\mathbb{E}M\approx\frac{1}{2}\int_\Omega Tf(\bm{x},t)\,d\bm{x}\,dt\,\propto T^{1-\beta}R^{-\gamma}\mathbb{E}N\,.
\label{Mheur}
\end{equation}
Fixing $\mathbb{E}N$ as a constant sets $R\propto   T^{-1/2}$, and then $\mathbb{E}M$ is independent of scale if and only if
\begin{equation}
\gamma=2(\beta-1)\,.
\end{equation}
We thus arrive at the general functional form for determining the zonal area of a given node:
\begin{equation}
f(\bm{x},t)=\alpha(\theta)t^{-\beta}r^{2-2\beta}\,.
\label{influ}
\end{equation}
Following (\ref{Mheur}), the total number of edges scales as $\mathbb{E}M\sim(\mathbb{E}N)^{2-\beta}$, therefore the networks generated are dense if $\beta<1$ and sparse if $\beta>1$. Notice also that the dependence on the radial coordinate $r$ means that zonal area expands with distance from the origin if $\beta<1$ and contracts when $\beta>1$. The ensembles of scale-invariant generated under this rule are therefore not generally translationally invariant in space. 

Translational invariance is, however, achieved by the special case $\beta=1, \alpha(\theta)\equiv \alpha$, that is $f(\bm{x},t)=\alpha/t$. For this choice the integral in (\ref{Mheur}) does not hold, so we must check more carefully. Since the maximum degree of a vertex is of course less than the number of nodes in the graph we can in this case compute
\begin{equation}
\begin{split}
\mathbb{E}M&\approx\frac{1}{2}\int_\Omega \min\{Tf(\bm{x},t),\mathbb{E}N\}\,d\bm{x}\,dt\\&=\frac{\alpha \pi R^2}{2}\int_{0}^{T}\min\{T/t,\mathbb{E}N\}\,dt\\&=\frac{\alpha \pi R^2T}{2}(1+\log(\mathbb{E}N)) \approx \frac{\alpha}{2}\mathbb{E}N\log(\mathbb{E}N)\,.
\end{split}
\end{equation} 
So again we find that for this choice of attachment rule the total number of edges in the resulting graph is independent of the spatial and temporal scale. Fig.~\ref{fig1} contrasts a classic RGG with an example of such a translationally and spatiotemporally scale-invariant geometric random graph.

\section{Degree distributions}\label{degs}

\begin{figure}[t]
(a)\includegraphics[width=100pt, trim=20 0 0 0, clip=true]{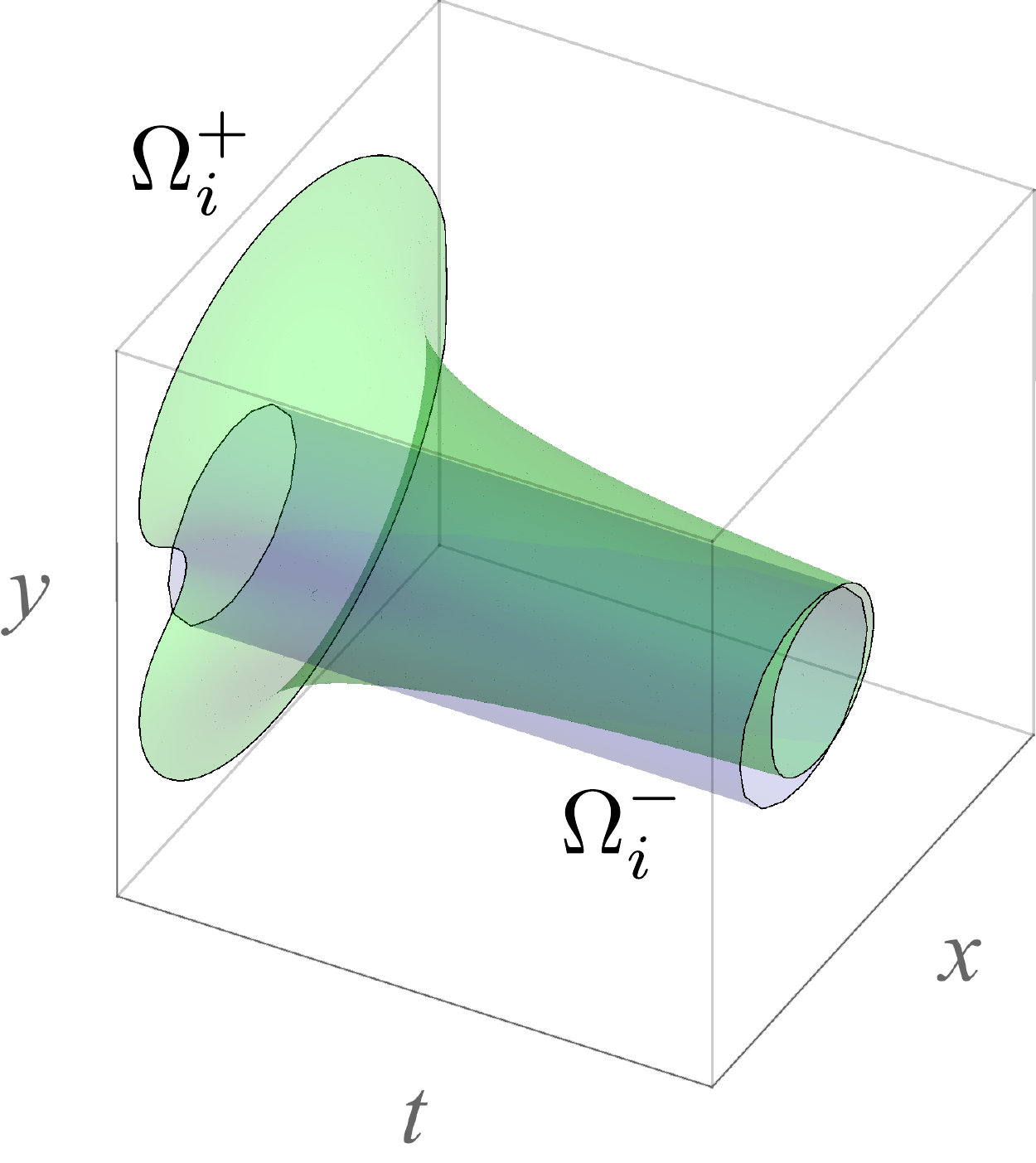}\hspace{20pt}(b)\includegraphics[width=100pt, trim=20 0 0 0, clip=true]{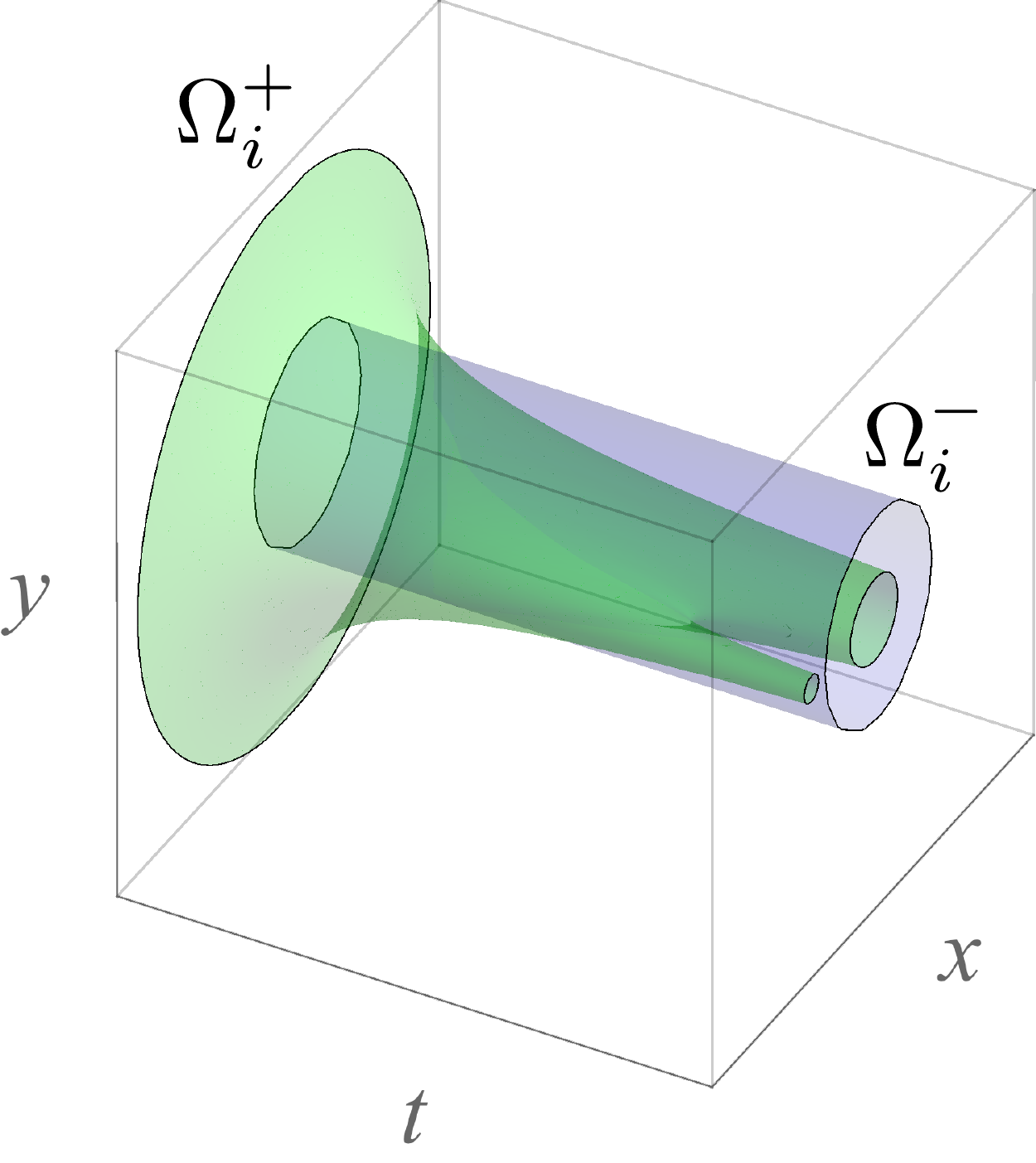}
\caption{Illustration of the regions $\Omega_i^-$ and $\Omega_i^+$, with $\beta=1/2,\, \bm{x}_i=(1.5,0)$ in (a) and $\beta=3/2,\, \bm{x}_i=(1,0)$ in (b).}
\label{fig2}
\end{figure}

Let us fix a node $i$ with co-ordinates $(\bm{x}_i,t_i)$, and examine the distribution of in-degree and out-degree. This node is the target of an inbound edge from every other node inside the cylindrical region $\Omega_i^-$ consisting of points whose spatial coordinates lie inside the zone of $i$. Conversely, node $i$ has a directed edge leading from it to every node $j$ for which $\bm{x}_i\in\Omega^-_j$ (see Fig.~\ref{fig2} for an illustration). Writing $\Omega^+_i$ for the region of possible locations of such nodes,
\begin{equation}
\begin{split}
&\Omega^-_i=\Big\{(\bm{x},t)\in \Omega:\pi|\bm{x}-\bm{x}_i|^2\leq f(\bm{x}_i,t_i)\Big\}\Bigg.\,,\\
&\Omega^+_i=\Big\{(\bm{x},t)\in \Omega:\pi|\bm{x}-\bm{x}_i|^2\leq f(\bm{x},t)\Big\}\,.
\end{split}
\end{equation}

The number of nodes that fall within either of these regions is a Poisson random variable with mean equal to the region volume. Therefore the (ensemble) probability that a node with known coordinates $(\bm{x}_i,t_i)$ has a given in- or out-degree is simply
\begin{equation}
\mathbb{P}(k_i^\pm=k)=\frac{\textrm{vol}(\Omega_i^\pm)^k \exp\big\{-\textrm{vol}(\Omega_i^\pm)\big\}}{k!}\,.
\label{poiss}
\end{equation}

The spatial nature of the connection mechanism we employ induces correlations between nodes, for the simple reason that if $i\to j$ and $j\to l$ it is more likely that $i$ and $l$ are spatially proximate, and hence $i\to l$. However, one might hope to obtain a reasonable approximation to the degree distribution of a realized network by simply integrating Eq.~(\ref{poiss}) over the uniform distribution of node location. For the in-degree $p_k^-$, this procedure works well. We compute
\begin{equation}
\mathbb{E}p_k^-=\frac{1}{\pi R^2Tk!}\int_\Omega e^{L_k(\bm{x},t)}\,d\bm{x}\,dt\,,
\label{pkm}
\end{equation}
where $L_k(\bm{x},t)=k\log \big(Tf(\bm{x},t)\big)- Tf(\bm{x},t)$. Laplace's method offers a considerable simplification. $L_k$ achieves its maximum in $t$ at
\begin{equation}
t_\star(\bm{x})=(\alpha(\theta)r^{2-2\beta}T/k)^{1/\beta}\,,
\end{equation}
expanding the exponent around this point, we find
\begin{equation}
L_k(\bm{x},t)\approx k\log(k)-k-\frac{k\beta^2}{2}\left(\frac{t}{t_\star(\bm{x})}-1\right)^2\,.
\end{equation}
Inserting this approximation into (\ref{pkm}) and integrating we find
\begin{equation}
\begin{split}
\mathbb{E}p_k^-&\approx \frac{k^ke^{-k}}{\pi R^2Tk!}\int_\Omega e^{-\frac{k\beta^2}{2}\left(t/t_\star(\bm{x})-1\right)^2}\,d\bm{x}\,dt\\
&\approx \left(\frac{k^ke^{-k}}{k^{1/2+1/\beta}k!}\right)\frac{\sqrt{2} \int_{-\pi}^\pi\int_0^R \alpha(\theta)^{1/\beta}r^{2/\beta-2}\,dr\,d\theta}{\beta \pi^{1/2}R^2T^{1-1/\beta}}\\
&\approx \left(\frac{1}{k^{1+1/\beta}}\right) \frac{\int_{-\pi}^\pi \alpha(\theta)^{1/\beta}\,d\theta}{\beta \pi^{1/\beta}}(\mathbb{E}N)^{1/\beta-1}\,.
\label{pki}
\end{split}
\end{equation}
where the second line follows from Gaussian integration on a time axis extended to $\pm\infty$, and the third from  using Stirling's approximation for the factorial. 
\begin{figure}[t]
\includegraphics[width=240pt, trim=0 0 0 0]{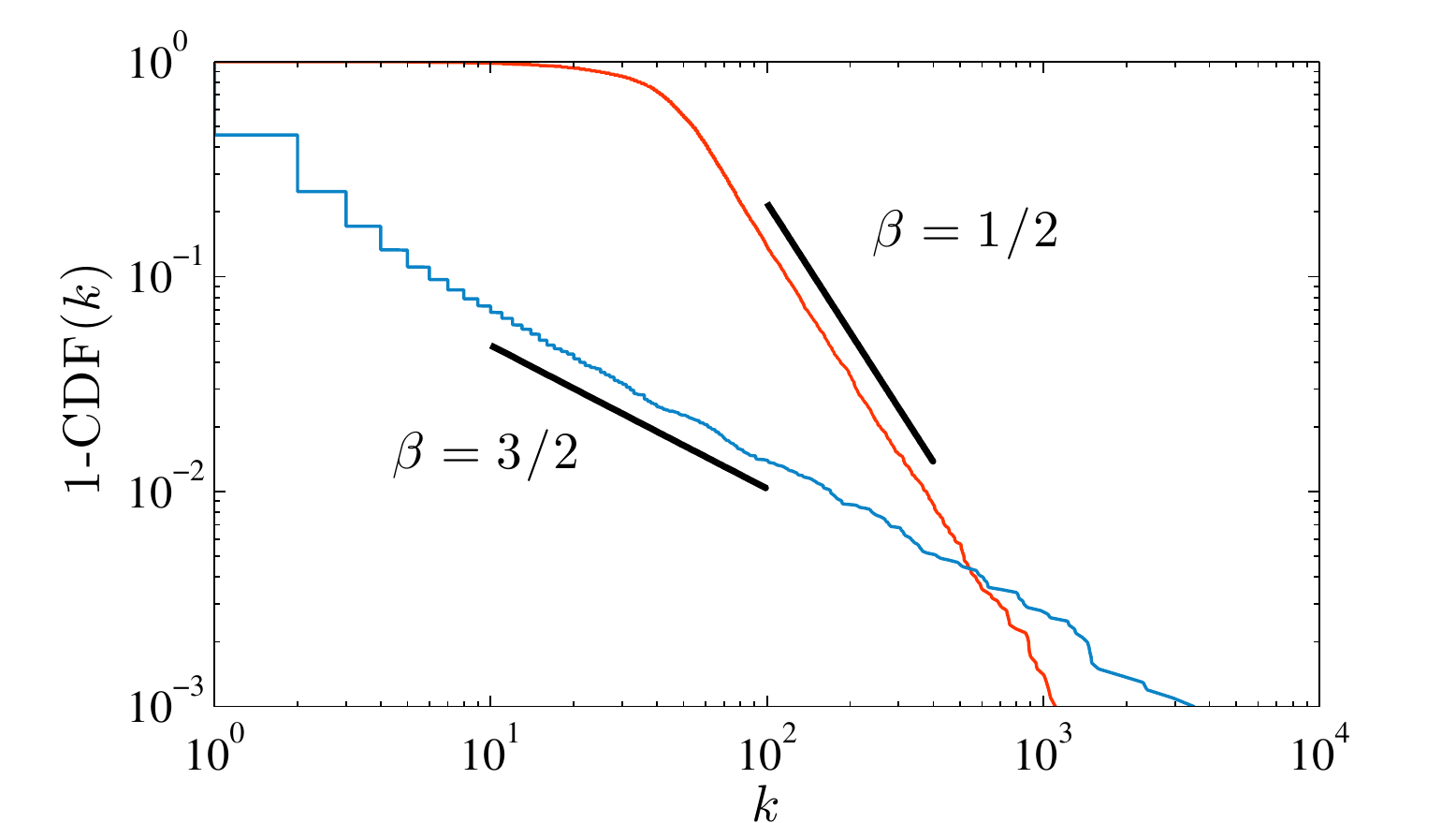}
\caption{The thin colored lines show the tail distribution of in-degree for scale-invariant random geometric graphs with $\beta=1/2$ (red) and $\beta=3/2$ (blue). In each case a single graph of size $N=10^4$ was generated and the empirical tail distribution $1-\text{CDF}(k)$ calculated. Black lines indicate the theoretical result $\mathbb{P}(k_i>k)\sim k^{(-1/\beta)}$ derived from the power law distribution (\ref{pki}).}
\label{fig_tails}
\end{figure}

This calculation becomes exact in the limit $k\to\infty$, meaning that the tail of the in-degree distribution has a power-law with exponent $-(1+1/\beta)$. One important caveat to note is that formally the integrals as they written in (\ref{pki}) may not converge, due to singularities at the origin in time ($\beta\leq1$) or space ($\beta>1$). Mathematically, this is simply an artifact of us not taking care with zones that may overlap the boundary of the sampling region $\Omega$, however, this observation is part of a wider phenomenon with major consequences for the distribution of out-degrees.

Examining (\ref{influ}), it is clear to see that nodes arriving very early in the process ($t\ll1$) enjoy a considerable advantage over more typical nodes, having extremely large zonal areas. When $\beta>1$, the same is true of nodes close to the center of the spatial domain. In fact, any node falling in the region
\begin{equation}
\Gamma=\left\{(r,\theta,t)\,:\,\alpha(\theta) t^{-\beta} r^{2(1-\beta)}\geq \pi R^2\right\}
\end{equation}
has a zone the same size as $\Omega$. In a particular network realization, the possible presence of such a `hub' node would have the effect of increasing the degree of almost every other node by one, essentially shifting the entire out-degree distribution one place to the right. The number of such hubs is itself a Poisson random variable, with mean 
\begin{equation}
\textrm{vol}(\Gamma)= \alpha^{1/\beta}\beta\pi^{1-1/\beta}\,.
\end{equation}
\begin{figure}[t]
\includegraphics[width=230pt, trim=80 325 80 320]{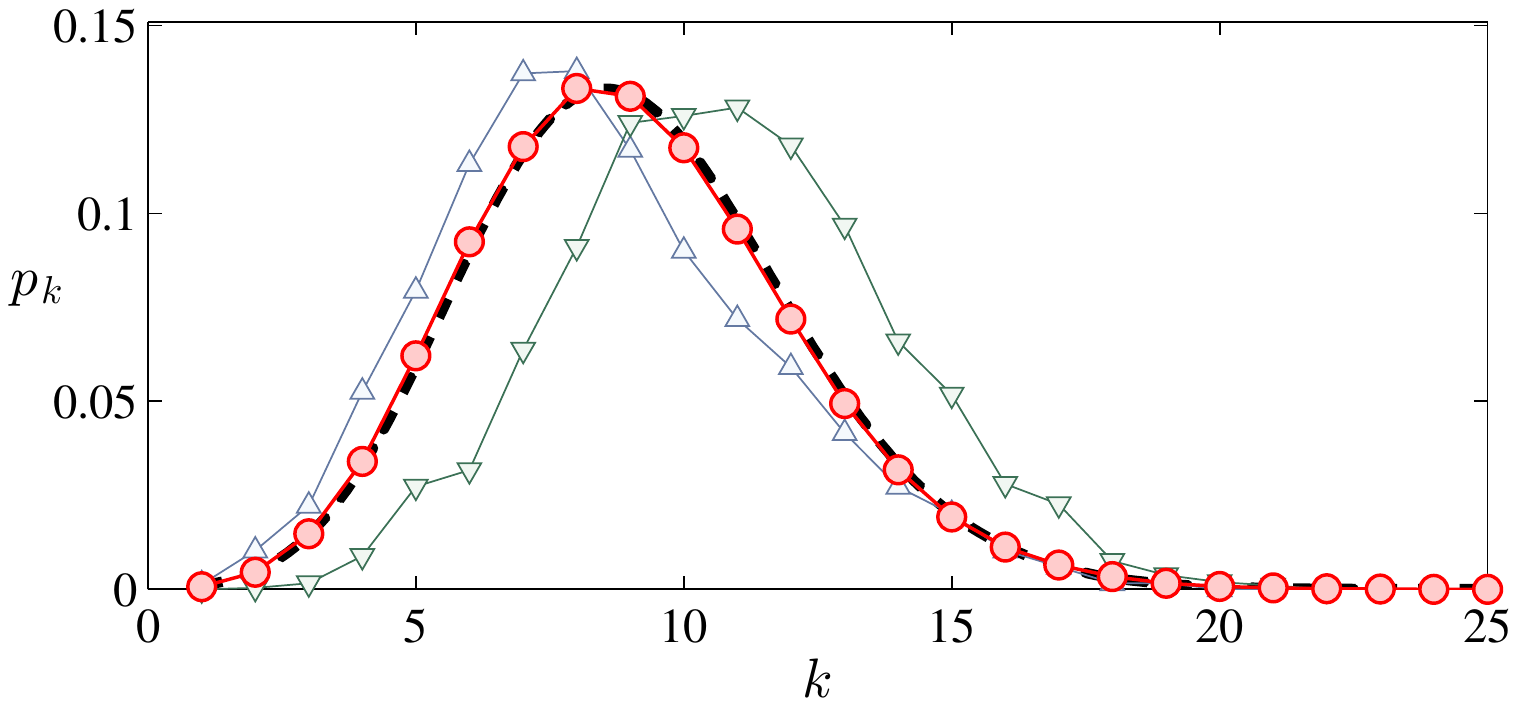}
\caption{Out degree distribution for the scale invariant geometric random graph with $\beta=1$, $\alpha(\theta)\equiv1$ and $\pi R^2T=10^3$. The red circles show the out degree distribution averaged over 100     realizations, while the black curve shows the result of equation (\ref{out_1}). Blue and green triangles are the empirical degree distributions from two of the 100 graphs sampled --- their horizontal shift results from the random number of hubs.}
\label{out_degree_fig}
\end{figure}

For nodes outside of the hub region in large networks ($T\gg1$) we can expect that the volume of $\Omega_i^+$ is dominated by the large $T$ tail, in which we can approximate the area segment $\pi|\bm{x}-\bm{x}_i|^2=f(\bm{x},t)$ by the circle $\pi|\bm{x}-\bm{x}_i|^2=f(\bm{x}_i,t)$. The result for $\beta\neq1$ is
\begin{equation}
\begin{split}
\text{vol}(\Omega_i^+\setminus \Gamma)&\approx \alpha(\theta_i)r_i^{2-2\beta}\int_{T_0(r_i,\theta_i)}^T \hspace{-6mm}t^{-\beta}	\,dt\\
&=(\mathbb{E}N)^{1-\beta}\frac{\alpha \pi^{\beta-1}}{(\beta-1)(\beta-2)}+\frac{\alpha^{1/\beta}\beta \pi^{1-1/\beta}}{\beta-1}\,,
\label{volnohub}
\end{split}
\end{equation}
where the lower limit $T_0(r,\theta)$ of the integral was the time boundary of the hub region $\Gamma$ along the line with fixed $r$ and $\theta$. The out-degree distribution of node $i$ is then given by a Poisson with mean $\text{vol}(\Omega_i^+\setminus \Gamma)$, right-shifted by a Poisson number of hubs with mean $\text{vol}(\Gamma)$. 

For the translationally invariant case $\beta=1$ and $\alpha(\theta)\equiv \alpha$ there is again a logarithmic correction. The hub region for this model is a simple cylinder with volume $\alpha$, but the integral in equation~(\ref{volnohub}) evaluates to $\text{vol}(\Omega_i^+\setminus \Gamma)= \alpha\log (\mathbb{E}N/\alpha)$. Averaging over the number $h$ of possible hubs, the ensemble out-degree distribution in this case is 
\begin{equation}
\begin{split}
\mathbb{E}p_k^+&=\sum_{h=0}^k\frac{\alpha^h e^{-\alpha}}{h!}\frac{(\alpha\log(\mathbb{E}N/\alpha))^{k-h} e^{-\alpha\log(\mathbb{E}N/\alpha)}}{(k-h)!} \\
&=\frac{1}{k!}\big(\alpha(1+\log(\mathbb{E}N/\alpha))\big)^ke^{-\alpha(1+\log(\mathbb{E}N/\alpha))}\,.
\end{split}
\label{out_1}
\end{equation}
Fig.~\ref{out_degree_fig} compares this prediction to numerical results. 

\section{Clustering and percolation}\label{cluperc}

Experimentally, we observe that the generated networks are typically highly clustered, primarily due to the spatial attachment mechanism. There is also an effect due to the scale free distribution of in-degree, which can bee seen by the following argument. Nodes with small in-degree typically have small zonal areas and thus nodes connecting to them are likely to be spatially close and hence connected to each other. The scale-free degree distribution of in-degree implies that there are very many more nodes of small in-degree than large, so they dominate the mean clustering coefficient. Note that this behaviour is quite different from that of other networks with scale-free distributions, see \textit{e.g.} \cite{Bollobas}. For the translationally invariant case $(\beta=1\,,\alpha(\theta)\equiv\alpha)$ we can go further than this heuristic argument, deriving a scaling relation between node degree and clustering. Suppose node $i$ has time-coordinate $t$ and node $j$ links to $i$. If $j$ has time-coordinate $s$ and we make the assumption that the overlap of the zones of $i$ and $j$ is approximately $\min\{\alpha/s,\alpha/t\}$ (this is justified if either of $s$ or $t$ is large, which is common in networks with $T\gg1$), then the probability that another node $\ell$ in the influence zone of $i$ is also in the influence zone of $j$ is $\min\{1,t/s\}$. Integrating over possible values of $s$, we find $C(t)\approx t(1+\log(T/t))/T$. For the case $\beta=1\,,\alpha(\theta)\equiv\alpha$ the mean degree of nodes with time-coordinate $t$ is simply $Tf(x,t)=\alpha T/t$, so together we find
\begin{equation}
C(k)\approx \alpha(1+\log(k/\alpha))/k\,.
\label{Ck}
\end{equation}

Another central question in the study of random networks is that of percolation, which has been thoroughly explored in the standard RGG models \cite{Penrose,Alon,Ziff}. The possible existence of infinite connected clusters is by definition a property of the system in the thermodynamic limit of large numbers of nodes.  We have numerically investigated the percolation transition in the scale invariant ensemble, for simplicity considering only the rotationally invariant case $\alpha(\theta)=\alpha N^{\beta-1}$. The additional scaling with $N$ here is required to keep the mean degree fixed as $N$ grows and hence obtain a meaningful thermodynamic limit. Writing $\rho_i$ for the fractional size of the strongly connected component containing node $i$, we compute the size of the largest component $\rho=\max_i\rho_i/N$, which increases from $N^{-1}$ to 1 as $\alpha$ varies in $[0,\infty)$.
\begin{figure}[t]
\begin{picture}(100,100)
\put(-70,0){\includegraphics[width=245pt, trim=25 0 40 0]{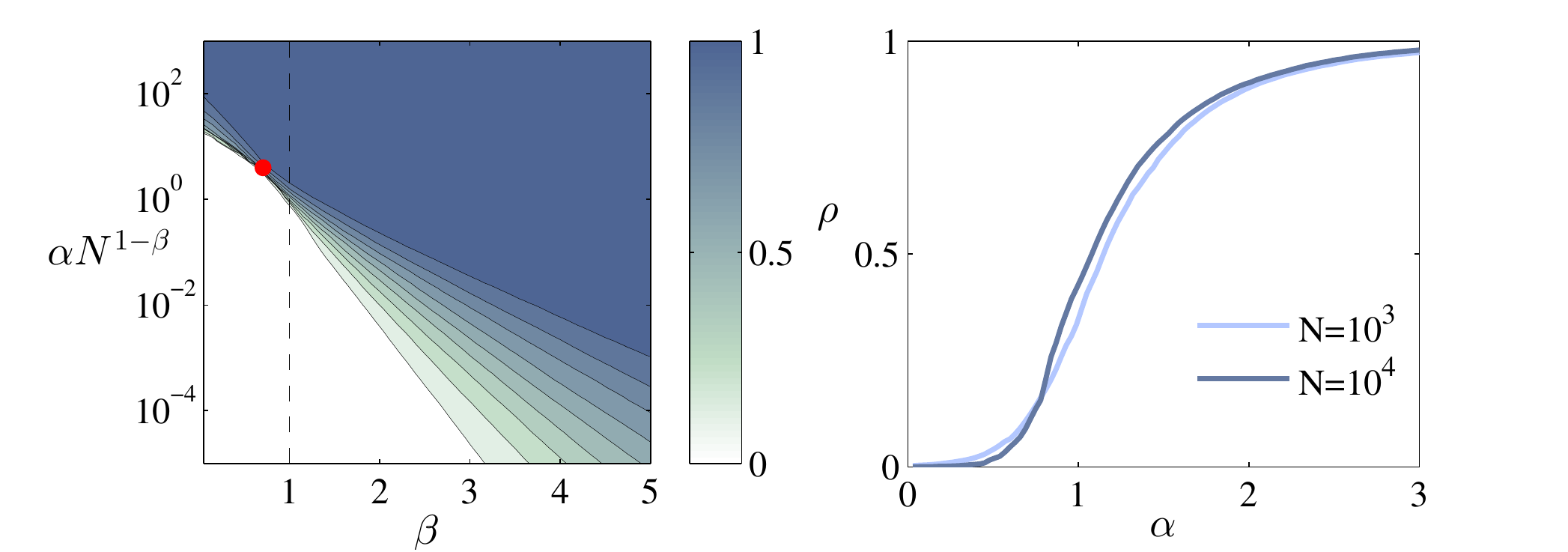}}
\put(-70,5){(a)}
\put(60,5){(b)}
\end{picture}
\caption{Percolation in the scale invariant RGG. (a) shades show the fractional size of the strongly connected component containing an average node in graphs generated from a fixed collection of $N=10^4$ uniformly distributed nodes, with various choices of exponent $\beta$ and connection length scale $\alpha(\theta)\equiv\alpha N^{1-\beta}$. (b) the $\beta=1$ slice for $N=10^2,10^3,\textrm{and}\,10^4$, averaged over $S=10^3,10^2,\textrm{and}\,10$ samples, respectively.}
\label{perc_fig}
\end{figure}
The numerical results shown in Fig.~\ref{perc_fig} present a several puzzles. Noting the logarithmic scale on the vertical axis of the left panel, we can tentatively suggest that for larger $\beta$ the model appears to percolate for arbitrarily small $\alpha$ in the limit $N\to\infty$. The inflection point of the level lines appears to be close to $(0.7,4)$, shown as a red dot, suggesting a critical point in this region. The behavior along the slice $\beta=1$, shown in the right panel, is also unexpected. Our earlier calculation showed that the mean degree is $1+\log(N)$ in this case, yet it appears that a logarithmic correction to $\alpha$ is not required to achieve data collapse for different $N$.

\section{Webgraph comparison}\label{web}

The properties of scale invariant geometric random graphs derived above are not of purely intellectual interest, but in fact show similarity to a particular class of real networks: web graphs. These are directed networks whose nodes correspond to web pages, and a directed link connects page $i$ to page $j$ if there exists a hyperlink on page $i$  referring to page $j$. Previous authors have proposed causal mechanisms relating to the `popularity' and `similarity' of pages \cite{Menczer,Papadopoulos} to explain the local connectivity properties of web graphs. Here, we show that essentially the same behaviour emerges as a consequence of the single simple property of scale invariance. We compare the web graph \emph{Web-Stanford} \cite{Leskovec} to our scale invariant model.

\begin{figure}[t]
\begin{picture}(100,100)
\put(-70,0){\includegraphics[width=240pt, trim=50 20 50 5]{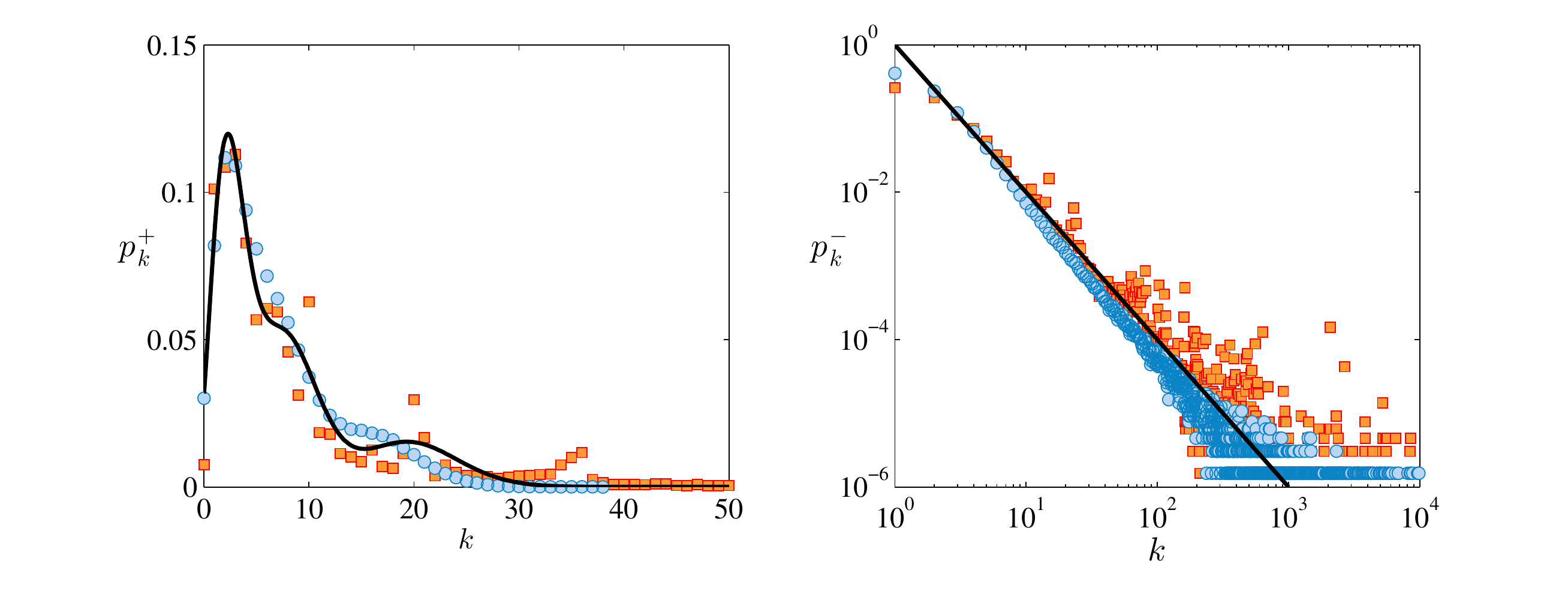}}
\put(-70,0){(a)}
\put(52,0){(b)}
\end{picture}
\caption{In-degree (a) and out-degree (b) distributions of a scale-invariant geometric random graph (blue circles) and the Stanford webgraph (orange squares), compared with analytical predictions of Eq.~(\ref{pki}) and Eq.~(\ref{out_1}) (black lines).}
\label{fig5}
\end{figure}

The Stanford webgraph exhibits a heavy-tailed in-degree distribution with exponent approximately $2$; we therefore must choose $\beta=1$ to fit our model. To fit the out-degree distribution, we are able to choose the angular dependence of influence zone $\alpha(\theta)$ as we wish. For simplicity we take a piecewise constant functional form whereby $\alpha(\theta)$ takes values $\{\alpha_1,\ldots \alpha_L\}$ on domains of size $\{2\pi w_1,\ldots,2\pi w_L\}$. Following the derivation of out-degree distribution from earlier we find a mixture Poisson 
\begin{equation}
\mathbb{E}p_k^+=\sum_l w_l e^{-c_l}c_l^k/k!\,,
\end{equation}
where $c_l=\alpha_l(1+\log(N/\alpha_l))$. A reasonable fit to the Stanford webgraph is achieved by the values $\{1.4, 0.55, 0.17\}$ and weights $\{0.17,0.36,0.47\}$. In principle we could choose a more complex form in order to achieve a closer fit, but we find three-component mixture to be sufficient for illustration purposes. From a modelling point of view the interpretation is that pages in the webgraph can be loosely grouped into three `types' that have different typical zones of influence. In Fig.~\ref{fig5} the in- and out-degree distributions of the Stanford webgraph is compared with that of a single randomly generated scale-invariant network, as well as our predictions for the ensemble average. 

\begin{figure}[t]
\includegraphics[width=240pt, trim=0 20 0 0]{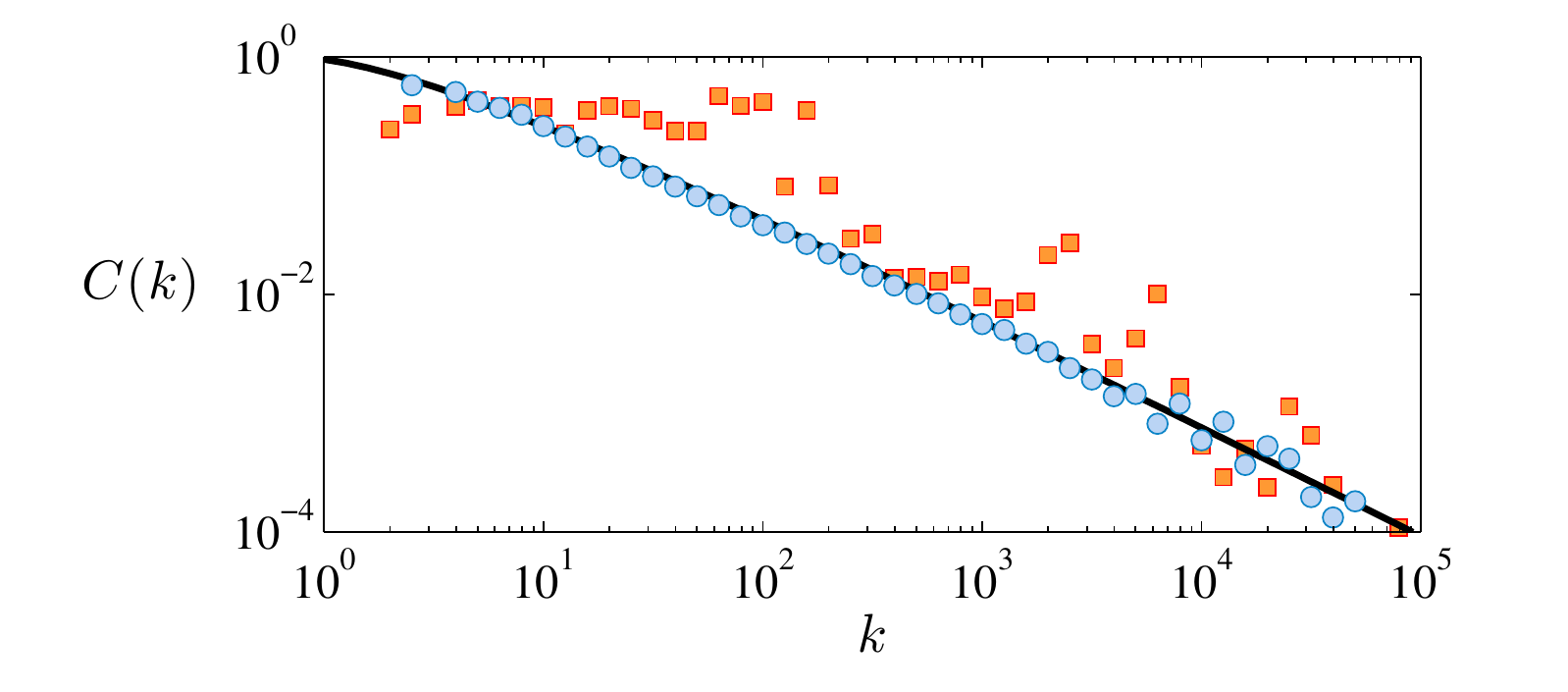}
\caption{Local clustering coefficient as a function of node in-degree in the scale-invariant geometric random graph (blue circles) and the Stanford webgraph (orange squares), compared with the theoretical prediction of Eq.~(\ref{Ck}). Note that the data has been binned on the $k$-axis to improve visibility (shown is the average of $C(k)$ for nodes with degree $\approx k$).}
\label{fig6}
\end{figure}

A simple test of the descriptive power of a fitted model of an empirical network is to as if it succeeds in predicting properties that were not explicitly fitted. In Fig.~\ref{fig6} we compare the correlation between degree and clustering coefficient for nodes in our fitted model and the webgraph. The agreement is by no means perfect, but we do observe in both graphs anti-correlation of approximately the same magnitude.


\section{Discussion}
The purpose of this article has been to initiate the study of a new class of geometric random graphs that are statistically invariant to changes of spatial and temporal scale. We have shown how this simple constraint leads to a graph ensemble that is governed by a single scale parameter $\beta$ and a function $\alpha(\theta)$ that determines local connectivity and spatial heterogeneity. Analytical arguments allow for the prediction of the degree distributions and degree-clustering correlations, which we have shown to provide a reasonable fit to those of the Stanford webgraph.

Our model is built on some of the same general ideas as explored in the ``cosmological'' RGGs of e.g. \cite{Krioukov}. One might therefore ask if these models are scale-invariant in the sense we use here. Certainly, they have a scale-free degree distribution (with $\beta=2$), however, the spatiotemporal scaling behaviour is quite different. Consider the RGG built on (1+1)-dimensional de Sitter space. Fix the desired number of nodes $N\approx \pi e^T\delta$, and scale the timespan by a factor $A>0$. Then the density $\delta_A$ should satisfy $ N\approx e^{AT}\delta_A$, i.e. $\delta_A\sim Ne^{-AT}$. In this case the mean degree behaves as $\langle k\rangle=\int^{AT}_0e^ss\delta_A\,ds/N\approx AT$, which clearly varies with the scaling transformation. Similarly the hyperbolic RGG of \cite{Krioukov1} is also not scale invariant with respect to the mean degree --- see equation (13) in that work. 

The notion of scale invariance is intimately related with that of self-similarity of networks, as studied previously in \cite{Serrano1,Serrano2}, for example. In particular, \cite{Serrano2} discusses an extremely general notion of self similarity in parameterized network ensembles. Our model can indeed be cast in this form, raising hopes of a theoretical analysis of the percolation transition. Unfortunately, none of the three specific types of self-similar networks discussed in \cite{Serrano2} apply directly to our model, so more work is need to make this link explicit.

The work we have presented has several limitations. In the model definition, we restricted ourselves to two-dimensional space and a single ``time'' axis that parametrizes zonal area. One possible direction for future studies would be to generalize to arbitrary dimensions, or to find a conceptually neat way to dispose of the time axis. On the subject of the calculations presented here, although their results agree well with simulations, most have relied on approximations that we have not specifically controlled. It would absolutely be desirable to formulate more rigorous statements. Finally, some readers may find the comparison to empirical webgraphs to be somewhat superficial, and they would be correct. If scale-invariant geometric random graphs are to find applications as, for example, null-models in empirical analysis then a much more nuanced and careful approach will be necessary. 

Several more exciting directions for future research have also been opened. The most important of these is to develop a precise theory percolation in scale-invariant graphs. Our numerical work shown in Fig.~\ref{perc_fig} suggests a critical point at a non-trivial coordinate in $(\alpha,\beta)$ space; clarification of this behavior would be very desirable. There is also the possibility that the study these random graphs may offer insight into general aspects of real-world networks. We have shown here how, for example, scale-free degree distributions can occur as a epiphenomenon of scale-invariance, which itself is a common property of critical systems. The ubiquity of scale-free networks in the real-world might therefore point to underlying generative mechanisms exhibiting self-organized criticality.

\section*{Acknowledgments}
TR acknowledges support from the Royal Society. ZX is supported by the National University of Defense Technology Graduate Teaching Reform Project No.201406-01.

\end{document}